\documentstyle[12pt]{article}
\newcommand{\bce}{\begin{center}}
\newcommand{\ece}{\end{center}}
\newcommand{\beq}{\begin{equation}}
\newcommand{\eeq}{\end{equation}}
\newcommand{\bea}{\vspace{0.25cm}\begin{eqnarray}}
\newcommand{\eea}{\end{eqnarray}}

\newcommand{\ba}{\begin{array}}
\newcommand{\ea}{\end{array}}

\newcommand{\doublespace}{
    \renewcommand{\baselinestretch}{1.6}\large\normalsize}

\def\lsim{\mathrel{\rlap{\lower4pt\hbox{\hskip1pt$\sim$}}
    \raise1pt\hbox{$<$}}}	  %less than or approx. symbol
\def\gsim{\mathrel{\rlap{\lower4pt\hbox{\hskip1pt$\sim$}}
    \raise1pt\hbox{$>$}}}	  %greater than or approx. symbol

\def\Pom{{\bf I\!P}}

\def\lsim{\mathrel{\rlap{\lower4pt\hbox{\hskip1pt$\sim$}}
    \raise1pt\hbox{$<$}}}         %less than or approx. symbol
\def\gsim{\mathrel{\rlap{\lower4pt\hbox{\hskip1pt$\sim$}}
    \raise1pt\hbox{$>$}}}         %greater than or approx. symbol

\def\Pom{{\bf I\!P}}

  \advance\oddsidemargin  by -1in
  \advance\evensidemargin by -1in
  \advance\topmargin      by -0.5in
\def\lsim{\mathrel{\rlap{\lower4pt\hbox{\hskip1pt$\sim$}}
    \raise1pt\hbox{$<$}}}         %less than or approx. symbol
\def\gsim{\mathrel{\rlap{\lower4pt\hbox{\hskip1pt$\sim$}}
    \raise1pt\hbox{$>$}}}         %greater than or approx. symbol

\def\Pom{{\bf I\!P}}
\def\beq{\begin{equation}}
\def\endeq{\end{equation}}
\def\arr{\begin{eqnarray}}
\def\endarr{\end{eqnarray}}
\makeindex

\doublespace

\begin{document}

%\large

%{ \huge The last update:  26 December 1993\vspace{2.0cm}\\}
\phantom{.}{\bf \Large \hspace{10.0cm} KFA-IKP(Th)-1994-1 \\
\phantom{.}\hspace{11.7cm}3 January 1994\\ }
%}

\begin{center}
{\bf\it \huge  The spectrum and solutions   of the generalized
BFKL equation   for total cross section}
\vspace{0.3cm}\\
{\bf \large
N.N.~Nikolaev$^{a,b}$, B.G.~Zakharov$^{a,b}$ and V.R.Zoller$^{a,c}$
\bigskip\\}
{\it
$^{a}$IKP(Theorie), KFA J{\"u}lich, 5170 J{\"u}lich, Germany
\medskip\\
$^{b}$L. D. Landau Institute for Theoretical Physics, GSP-1,
117940, \\
ul. Kosygina 2, Moscow 117334, Russia.\medskip\\
$^{c}$ Institute for Theoretical and Experimental Physics,\\
Bolshaya Cheremushkinskaya 25, 117259 Moscow, Russia.
\vspace{0.5cm}\\ }
{\Large \bf
Abstract}\\
\end{center}
The colour dipole cross section is the principal
quantity in the lightcone $s$-channel description of the
diffractive scattering. Recently we have shown that the dipole
cross section satisfies the generalized BFKL equation. In this
paper we discuss properties and solutions of our generalized BFKL
equation with allowance for the finite
gluon correlation radius $R_{c}$. The latter is introduced
in a   gauge invariant manner. We present estimates of the
intercept of the pomeron and find the asymptotic form of the
dipole cross section.
 \smallskip\\
         \begin{center}
{\large Submitted to {\sl Physics Letters B}\\
E-mail: kph154@zam001.zam.kfa-juelich.de}
\end{center}

%\doublespace
\pagebreak

%-----------------

%                 Section  1

%-----------------

\section{Introduction.}

%-----------------

The asymptotic behavior of high-energy scattering in perturbative
QCD is usually discussed in terms of the Balitskii-Fadin-Kuraev-Lipatov
(BFKL) equation for the differential gluon distribution function [1-3].
In this communication we discuss properties of the
 generalized BFKL equation
directly for the total cross section, which was
proposed by us recently
[4,5]. Our equation enables one
to introduce the finite correlation radius for the perturbative
gluons in a gauge invariant manner.
The starting point
of its derivation is the technique of multiparton lightcone
wave functions, developed by two of the present authors [4,6].
The principal observation is that the
transverse separations $\vec{\rho}_{i}$ and
the lightcone momentum partitions $z_{i}$ of partons in the
many-body Fock state are conserved in the scattering
process. Interaction of the $(n+2)$-parton Fock state is
described by the lightcone wave function
$\Psi_{n+2}(\vec{\rho}_{n+2},z_{n+2},...,\vec{\rho}_{1},z_{1})$
and the $(n+2)$-parton cross section
$\sigma_{n+2}(\vec{\rho}_{n+2},...,\vec{\rho}_{1})$, which are
perturbatively calculable [4] (here $n$ referes to the number
of gluons in the Fock state).

To the lowest order in the perturbative  QCD  one starts
with the $q\bar{q}$ Fock states of mesons ($qqq$ state for the
baryons) and with the scattering of the ($A$)
projectile and ($B$) target colour dipoles of transverse size
$\vec{r}_{A}$ and $\vec{r}_{B}$ (here $\vec{r}_{A},\vec{r}_{B}$ are
the two-dimensional vectors in the impact parameter plane)
\beq
\sigma_{0}(\vec{r}_{A},\vec{r}_{B})=
{32 \over 9}
\int {d^{2}\vec{k}
\over(k^{2}+\mu_{G}^{2})^{2} }
\alpha_{S}^{2}
\left[1-\exp(-i\vec{k}\vec{r}_{A})\right]
\left[1-\exp(i\vec{k}\vec{r}_{B})\right]  \, .
\label{eq:1.1}
\endeq
Here $\alpha_{S}(k^{2})$ is the running strong
strong coupling and in the integrand of (\ref{eq:1.1})
$\alpha_{S}^{2}$ must be understood as
$\alpha_{S}({\rm max}\{k^{2},{C^{2}\over r_{A}^{2}}\})
\alpha_{S}({\rm max}\{k^{2},{C^{2}\over r_{B}^{2}}\})$, where
$C\approx 1.5$ and we impose the infrared freezing of
$\alpha_{S}(k^{2})$ at large distances ([6] and see below).
Introduction
of the correlation radius for gluons $R_{c}=1/\mu_{G}$ is
discussed below.
In terms of the dipole-dipole cross section (\ref{eq:1.1}), the
perturbative part of the total cross section for the interaction
of mesons $A$ and $B$ equals
\beq
\sigma^{(pt)}(AB)=\langle \langle
\sigma(\vec{r}_{A},\vec{r}_{B}) \rangle_{A}
\rangle_{B}=
\int dz_{A} d^{2}\vec{r}_{A}
dz_{B} d^{2}\vec{r}_{B}
|\Psi(z_{A},\vec{r}_{A})|^{2}
|\Psi(z_{B},\vec{r}_{B})|^{2}
\sigma_{0}(\vec{r}_{A},\vec{r}_{B}) \,.
\label{eq:1.2}
\endeq
The irrefutable advantage of the representation
(\ref{eq:1.2}) is that it makes
full use of the {\sl exact}
diagonalization of the scattering matrix in
the dipole-size representation.
(Hereafter we discuss $\sigma_{0}(\vec{r}_{A},\vec{r}_{B})$
averaged over
the relative orientation of dipoles, as it appears in
Eq.~(\ref{eq:1.2}).)

The increase in the perturbative component of the
total cross section
comes from the rising muliplicity of perturbative
gluons in hadrons, $n_{g} \propto \log s$, times
$\Delta\sigma_{g}$ - the change in the dipole cross
section for the presence of gluons:
$\Delta\sigma^{(pt)} \sim n_{g}\Delta\sigma_{g}$ (here $s$ is
a square of the c.m.s energy).
The $n$-gluon Fock
components of the meson give contributions $\propto \log^{n}s$
to the total cross section.

The peresentation is organized as follows. We briefly review the
derivation of our generalized BFKL equation for the dipole cross
section
and discuss its BFKL scaling limit. Then we discuss the impact of
the finite correlation length for gluons $R_{c}$
and of the running QCD
coupling on the spectrum of our generalized BFKL equation. Our
principal conclusion is that the generalized BFKL kernel has
the continuous spectrum, so that the partial waves of the
scattering amplitude have a cut in the complex angular momentum
plane. We find the form of the dipole cross section for the
rightmost singularity in the $j$-plane and its intercept as a
function of the gluon correlation radius $R_{c}$. We conlcude
with comments on the admissible form of the boundary conditions
for the generalized BFKL equation and on restoration of
the factorization of asymptotic total cross sections.

%---------------------------------------
%                            Section 2
\section{Generalized BFKL equation for total cross section}

Now we sketch the derivation of our generalized BFKL equation [4,5]
for the dipole cross section.
The perturbative $q\bar{q}g$ Fock state generated
radiatively from the parent colour singlet $q\bar{q}$ state of
size $\vec{r}$ has the interaction cross section [4]
\beq
\sigma_{3}(r,\rho_{1},\rho_{2})=
{9 \over 8}[\sigma_{0}(\rho_{1})+\sigma_{0}(\rho_{2})] -
{1 \over 8}\sigma_{0}(r) \, ,
\label{eq:2.1}
\endeq
where $\vec{\rho}_{1,2}$ are separations of the
gluon from the quark and antiquark respectively, $\vec{\rho}_{2}=
\vec{\rho}_{1}-\vec{r}$. (Hereafter we suppress the target
variable $r_{B}$ and for the sake of brevity use $r=r_{A}$.)
The cross section $
\sigma_{3}(r,\rho_{1},\rho_{2})$ has gauge invariance
properties of $\sigma_{3}(r,0,r)=\sigma_{3}(r,r,0)=\sigma_{0}(r)$
and $\sigma_{3}(0,\rho,\rho)={9\over 4}\sigma_{0}(\rho)$. The
former shows that when the gluon is sitting on top of the (anti)quark,
the $qg (\bar{q}g)$ system is indistinguishable from the (anti)quark.
The latter shows that the colour-octet $q\bar{q}$ system of the
vanishing size is indistinguishable from the gluon, and ${9\over 4}$
is the familiar ratio of the octet and triplet couplings.
The increase of the cross section for the presence of
gluons equals
\beq
\Delta \sigma_{g}(r,\rho_{1},\rho_{2})=
\sigma_{3}(r,\rho_{1},\rho_{2}) -
\sigma_{0}(r)=
{9 \over 8}[\sigma_{0}(\rho_{1})+
\sigma_{0}(\rho_{2})-\sigma_{0}(r)]   \, \, ,
\label{eq:2.2}
\endeq
The lightcone density of soft, $z_{g}\ll 1$,
gluons in the $q\bar{q}g$ state derived
in [4] equals
\arr
|\Phi_{1}(\vec{r},\vec{\rho}_{1} ,\vec{\rho}_{2},z_{g})|^{2}=
{1 \over z_{g}} {1 \over 3\pi^{3}}
\mu_{G}^{2}
\left|g_{S}(r_{1}^{(min)})
K_{1}(\mu_{G}\rho_{1}){\vec{\rho}_{1}\over \rho_{1}}
-g_{S}(r_{2}^{(min)})
K_{1}(\mu_{G}\rho_{2}){\vec{\rho}_{2} \over \rho_{2}}\right|^{2} \, .
\label{eq:2.3}
\endarr
Here $g_{S}(r)$ is the effective colour charge,
$\alpha_{S}(r)=g_{S}(r)^{2}/4\pi$ is the running strong coupling,
$r_{1,2}^{(min)}={\rm min}\{r,\rho_{1,2}\}$, $K_{1}(x)$ is the
modified Bessel function,
$z_{g}$ is a fraction of the (lightcone) momentum of
$q\bar{q}$ pair carried by the gluon, and $\int dz_{g}/z_{g}
=log(s/s_{0})=\xi$ gives the usual logarithmic multiplicity of
radiative gluons. The wave function (\ref{eq:2.3}) counts only
the physical, transverse, gluons.

If $n_{g}(r)$ is the number of perturbative gluons in the
dipole $\vec{r}$,
\beq
n_{g}(r)=
\int
dz_{g}\,d^{2}\vec{\rho}_{1}\,\,
|\Phi_{1}(\vec{r},\vec{\rho}_{1} ,\vec{\rho}_{2},z_{g})|^{2}\, ,
\label{eq:2.4}
\endeq
then the weight of the
radiationless $q\bar{q}$ component will be renormalized
by the factor $[1-n_{g}(z,\vec{r})]$. As a result, the total
cross section with allowance for the perturbative gluons in the
beam dipole $A$ takes the form
\arr
\sigma(\xi,r)=
[1-n_{g}(z,\vec{r})]\sigma_{0}(r)+
\int dz_{g}\vec{\rho}\,\,
|\Phi_{1}(\vec{r},\vec{\rho}_{1} ,\vec{\rho}_{2},z_{g})|^{2}
\sigma_{3}(r,\rho_{1},\rho)_{2}]~~~~~~~~\nonumber\\
=
\sigma_{0}(r)+
\int dz_{g}\vec{\rho}_{1}\,\,
|\Phi_{1}(\vec{r},\vec{\rho}_{1} ,\vec{\rho}_{2},z_{g})|^{2}
\Delta\sigma_{g}(r,\rho_{1},\rho_{2})=\sigma_{0}(r)+\sigma_{1}(r)\xi
=[1+\xi{\cal K}\otimes]\sigma_{0}(r)
 \,,
\label{eq:2.5}
\endarr
where the kernel ${\cal K}$ is defined by [4,5]
\arr
\sigma_{1}(r)={\cal K}\otimes\sigma_{0}(r)=~~~~~~~~~~~~~~~~~~~~~~~~~~~~
{}~~~~~~~~~~~~~~~~~\nonumber\\
{3 \over 8\pi^{3}} \int d^{2}\vec{\rho}_{1}\,\,
\mu_{G}^{2}
\left|g_{S}(r_{1}^{(min)})
K_{1}(\mu_{G}\rho_{1}){\vec{\rho}_{1}\over \rho_{1}}
-g_{S}(r_{2}^{(min)})
K_{1}(\mu_{G}\rho_{2}){\vec{\rho}_{2} \over \rho_{2}}\right|^{2}
[\sigma_{0}(\rho_{1})+
\sigma_{0}(\rho_{2})-\sigma_{0}(r)]   \, \, .
\label{eq:2.6}
\endarr
Eq.~(\ref{eq:2.5}) shows that the effect of gluons can be reabsorbed
into the generalized dipole cross section $\sigma(\xi,r)$.
To higher orders in $\xi$, this generalized dipole cross section
can be expanded as
$
\sigma(\xi,r)=\sum_{n=0}{1\over n!}\sigma_{n}(r)
\xi^{n} $. Then, Eq.~(\ref{eq:2.6}) generalizes to
$\sigma_{n+1}={\cal K}\otimes \sigma_{n}$, and
\beq
{\partial \sigma(\xi,r) \over \partial \xi} ={\cal K}\otimes
\sigma(\xi,r)
\label{eq:2.7}
\endeq
is our generalized BFKL equation for the dipole
cross section.

Colour gauge invariance of the presented formalism is noteworthy.
Firstly, the dipole-dipole cross section $\sigma_{0}(r,R)$
vanishes at $r \rightarrow 0$ or $R\rightarrow 0$, because
gluons decouple from the colour-singlet state of vanishing size.
Secondly, for the same reason the wave function (\ref{eq:2.3})
vanishes at $r\rightarrow 0$.
Thirdly,
$\Delta \sigma_{g}(r,\rho_{1},\rho_{2})\rightarrow 0$ when
$\rho_{1}\rightarrow 0$ (or $\rho_{2}\rightarrow 0$), since by
colour charge conservation the quark-gluon system with the gluon
sitting on top of the (anti)quark is indistinguishable from the
(anti)quark, and the interaction properties of such a $q\bar{q}g$
state are identical to that of the $q\bar{q}$ state.
Therefore, the above introduction of a finite
correlation radius for gluons $R_{c}=1/\mu_{G}$, which
takes care of the perturbatie gluons not propagating beyond the
correlation radius $R_{c}$, is perfectly consistent
with gauge invariance constraints.

The renormalization of the weight of the radiationless $q\bar{q}$
Fock state in Eq.~(\ref{eq:2.5})
in a simple and intuitively appealing form takes care of the
virtual radiative corrections (in the BFKL formalism [1,2]
these very radiative corrections
are responsible for the reggeization of gluons).
The generalized BFKL equation (\ref{eq:2.6}) is both infrared and
ultraviolet finite.

%--------------------------------------------------------

%--------------------     Section 3

\section{The BFKL scaling limit}

In the BFKL scaling limit of $r, \rho_{1},\rho_{2} \ll R_{c}$
and fixed $\alpha_{S}$
\beq
\mu_{G}^{2}\left|K_{1}(\mu_{G}\rho_{1}){\vec{\rho}_{1}\over \rho_{1}}
-K_{1}(\mu_{G}\rho_{2}){\vec{\rho}_{2}\over\rho_{2}}\right|^{2} =
{r^{2} \over \rho_{1}^{2}\rho_{2}^{2}}   \, ,
\label{eq:3.1}
\endeq
the kernel ${\cal K}$ becomes independent of the gluon correlation
radius $R_{c}$ and with the fixed $\alpha_{S}$ it takes on the
scale-invariant form. The corresponding
eigenfunctions of Eq.~(\ref{eq:2.7}) are
\beq
E(\omega,\xi,r)=(r^{2})^{{1\over 2}+\omega}
\exp[\Delta(\omega)\xi]
\label{eq:3.2}
\endeq
with the eigenvalue (intercept)
[here $\vec{r}=r\vec{n}$,  $\vec{\rho}_{1}=r\vec{x}$  and
$\vec{\rho}_{2}=r(\vec{x}+\vec{n})$]
\arr
\Delta(\omega) = \lim_{\epsilon \to 0}
{3\alpha_{S} \over 2\pi^{2}}
\int d^{2}\vec{x}~~{2(\vec{x}^{2})^{{1\over 2}+\omega}-1
\over [\vec{x}^{2} +\epsilon^{2}]
[(\vec{x}+\vec{n})^{2}+\epsilon^{2}]}=
{3\alpha_{S} \over \pi}
\int_{0}^{1} dz
{z^{{1\over 2}-\omega}+z^{{1\over 2}+\omega}-2z \over z(1-z)}
\nonumber\\
=
{3\alpha_{S} \over \pi} [2\Psi(1)-
\Psi({1\over 2}-\omega)-\Psi({1\over 2}+\omega)]\, . ~~~~~~~~~~~~~~
\label{eq:3.3}
\endarr
Here $\Psi(x)$ is the digamma function, and
we have indicated the regularization which preserves the
symmetry of the kernel ${\cal K}$.
The final result for $\Delta(\omega)$ coincides with eigenvalues of
the BFKL equation [1-3].
 In the complex $j$-plane, the rightmost singularity
is located at $j=\alpha_{\Pom}=1+\Delta_{\Pom}$, where
\beq
\Delta_{\Pom}=\Delta(0)=
{12\log2\over \pi}\alpha_{S} \, .
\label{eq:3.4}
\endeq

When $\omega$ is real and varies from $-{1\over 2}$ to
0 and to ${1\over 2}$, also the intercept
$\Delta(\omega)$ is real and varies from $+\infty$ down to
$\Delta(0)=\Delta_{\Pom}$ and back to $+\infty$, along the cut
from $j=1+\Delta_{\Pom}$ to $+\infty$ in
the complex angular momentum $j$ plane. If $\omega =i\nu$ and $\nu$
varies from $-\infty$ to
0 and to $+\infty$, then the intercept
$\Delta(i\nu)$ is again real and varies from $-\infty$ up to
$\Delta(0)=\Delta_{\Pom}$ and back to $-\infty$, along the cut
from $j=-\infty$ to $j=1+\Delta_{\Pom}$
in the complex $j$-plane. The choice of the latter cut
is appropriate for the Regge asymptotics at $\xi \gg 1$ and
the counterpart of the conventional Mellin representation is
\beq
\sigma(\xi,r)=
\int_{-\infty}^{+\infty} d\nu \,f(\nu)E(i\nu,r,\xi) =
r\int {dr'\over (r')^{2}}K(\xi,r,r')\sigma(\xi=0,r')\, ,
\label{eq:3.5}
\endeq
where in the BFKL regime the evolution kernel (Green's function)
equals
\beq
K(\xi,r,r')={1\over \pi}\int d\nu \exp\left[2i\nu\log{r\over r'}
\right]\exp[\xi \Delta(i\nu)]
\propto {\exp(\Delta_{\Pom}\xi) \over \sqrt{\xi}}
\exp\left(-2{(\log r-\log r')^{2}\over
\xi \Delta^{''}(0)}\right) \, .
\label{eq:3.6}
\endeq
Here we have used the fact that $\Delta(i\nu)$ has the maximum
at $\nu=0$, so that the dominant at large $\xi$
saddle-point contribution
in (\ref{eq:3.6}) is evaluated using the expansion
\beq
\Delta(i\nu)=\Delta_{\Pom}-{1\over 2}\Delta''(0)\nu^{2}
\label{eq:3.7}
\endeq
Since the BFKL eigenfunctions (\ref{eq:3.2})
are oscillating functions of $r$, {\sl\'a priori}
it is not obvious that an arbitrary solution $\sigma(\xi,r)$
will be
positive-valued at all $\xi$ and $r$. The BFKL
kernel $K(\xi,r,r')$ of Eq.~(\ref{eq:3.6}) is manifestly
positive-valued at
large $\xi$, so that starting with the positive-valued
$\sigma(0,r)$ one obtaines the positive-valued asymptotical cross
section $\sigma(\xi,r)$.
In the BFKL regime, the rightmost $j$-plane singularity
corresponds to the asymptotic dipole cross section
$\sigma_{\Pom}(\xi,r) =
\sigma_{\Pom}(r)\exp(\xi\Delta_{\Pom})$, where
\beq
\sigma_{\Pom}(r)
\propto r \, .
\label{eq:3.8}
\endeq

More direct correspondance between our equations
(\ref{eq:2.6},\ref{eq:2.7}) in the scaling limit of
$\mu_{G}\rightarrow 0$ and the original
BFKL equation can be established
if one rewrites our Eqs.~(\ref{eq:2.6},\ref{eq:2.7})
as an equation for
the function
\beq
g(\xi,r)={3 \sigma(\xi,r)\over \pi^{2}r^{2}\alpha_{S}}\,
\label{eq:3.9}
\endeq
which in the BFKL scaling limit is simply the density of gluons
$g(x,k^{2})$
at the Bjorken variable $x=x_{0}\exp(-\xi)$ and the virtuality
$k^{2} \sim 1/r^{2}$, where $x_{0}\sim 0.1$-$0.01$ corresponds
to the onset of the leading-$\log({1\over x})$ approximation
 (the relation (\ref{eq:3.9}) holds for the
running coupling too, for the detailed derivation see [4,7]).
Indeed, making use of the conformal symmetry of the
fixed-$\alpha_{S}$ BFKL
equation [3], $\vec{r}$ can easily be traded for $\vec{k}$
and $\sigma(\xi,r)$ can be traded for $g(\xi,k^{2})$, and after
introduction of the function $\phi(\xi,k^{2})=g(\xi,k^{2})/k^{2}$
the original form of the BFKL equation
will be recovered from Eq.~(\ref{eq:2.6}) (in the BFKL
scaling limit
$\phi(\xi,k^{2})$ satisfies the same equation as the more often
considered $\partial g(\xi,k^{2})/\partial k^{2}$).

The GLDAP limit of Eqs.~(\ref{eq:2.6},\ref{eq:2.7}), i.e., the
limit of large $\log({1\over r^{2}})$ at finite $\xi$, was discussed
in great detail in [4], here we concentrate on $\xi \rightarrow
\infty$.

%-----------------------------------------

%----------------------------  Section 4

\section{The spectrum of the generalized
BFKL equation}

The `diffusion' kernel (\ref{eq:3.6}) makes it obvious that
the BFKL scaling approximation is not self-consistent in the
realm of realistic QCD:
starting with $\sigma(\xi=0,r)$ which was concentrated at the
perturbative small $r \lsim R \ll R_{c}$ one ends up at large
$\xi$ with $\sigma(\xi,r)$ which extends up to the
nonperturbative $r\sim R\exp(\sqrt{\xi\Delta''(0)}) > R_{c}$ .
Therefore, introduction of a certain infrared regularization
in the form of a finite gluon correlation length $R_{c}$ is
inevitable, and hereafter we concentrate on effects of finite
$R_{c}$ on the intercept $\Delta_{\Pom}$.
As we have emphasized above, our kernel ${\cal K}$ introduces
$R_{c}$ in the manifestly gauge invariant manner.
Interpretation of $\mu_{G}=1/R_{c}$ as an effective
mass of the gluon suggests the infrared freezing of the strong
coupling $\alpha_{S}(r)$ at $r> R_{f} \sim R_{c}$.
At $r\leq R_{f}$ we use
\beq
\alpha_{S}(r)={4\pi \over
\beta_{0}\log\left(
{C^{2}\over \Lambda_{QCD}^{2}r^{2}}\right)} \, ,
\label{eq:4.1}
\endeq
where $\beta_{0}=11-{2\over 3}N_{f} =9$ for $N_{f}=3$ active flavours,
we take $\Lambda_{QCD}=0.3\,$GeV and $C=1.5$ (for the discussion of
the scale factor $C$ see [6]).
At large $r>R_{f}=0.42$f we impose the simplest freezing
$
\alpha_{S}(r>R_{f})=\alpha_{S}^{(fr)}= 0.8 \, .
$
More sophisticated smooth freezing can easily be considered, but
it will be obvious that our principal conclusions do not depend
on the form of the freezing. The correlation radius $R_{c}$ and the
freezing coupling provide the minimal infrared regularization
of the perturbaton theory, and in the sequel we consider the
so-regularized generalized BFKL equation (\ref{eq:2.6})
as applicable at both small and large
radii $r$.

Although only the case of $R_{c}\sim    R_{f}$ is of the physical
interest, a study of (albeit unphysical)
limiting cases $R_{f}\rightarrow 0$ at finite $R_{c}$, and of
finite $R_{f}$ at $R_{c}\rightarrow \infty$,
is instructive for the insight into
how the spectrum of the $j$-plane singularities is modified by
the infrared regularization. The useful
observation is that in view of
Eqs.~(\ref{eq:3.5},\ref{eq:3.6}) the large-$\xi$ behavior of
solutions of the BFKL limit of
 Eq.~(\ref{eq:2.7}) is similar to that of solutions of the
'Schr\"odinger' equation
\beq
\left[-{\Delta''(0)\over 2}{\partial^{2} \over \partial z^{2} }
+V(z)\right]\Phi= -{\partial \over \partial \xi} \Phi
=\epsilon \Phi
\label{eq:4.2}
\endeq
for a particle of mass  $M=1/\Delta''(0)$ in the potential
$V(z)=-\Delta(0)$. Here the coordinate $z=\log r^{2}$,
the 'wavefunction' $\Phi(\xi,z)=\sigma(\xi,r)/r$ and the
intercept equals the 'energy' $\epsilon$
taken with the minus sign.

The first limiting case of $R_{f}\rightarrow 0$ corresponds to
introduction of a  finite gluon correlation radius $R_{c}$
at fixed $\alpha_{S}$. In this limiting case the intercept
$\Delta_{\Pom}$ will still be given by the BFKL formula (\ref{eq:3.4}).
Indeed, on the infinite semiaxis $\log r < \log R_{c}$
the kernel ${\cal K}$ retains its scaling properties,
the corresponding eigenfunctions  will be essentially identical to
the set (\ref{eq:3.2}), the spectrum of eigenvalues will be continuous
and the cut in the $j$-plane will be the same as
at $R_{c} \rightarrow \infty$. This is particularly obvious from
Eq.~(\ref{eq:4.2}), since on the the semiaxis $z<z_{c}=\log R_{c}^{2}$
the potential $V(z)=-\Delta_{\Pom}$ is flat, and
the corresponding Schr\"odinger operator
has the continuum spectrum starting with the minimal energy
$\epsilon=-\Delta_{\Pom}$. Evidently, this property of the spectrum
does not depend on the details of how the gluon correlation length
$R_{c}$ is introduced.
Similar conclusion on the spectrum of the infrared-cutoff BFKL
equation in the momentum-representation was reached in [8] in a model
with very different infrared cutoff.

The behaviour of solutions at large $r\gg R_{c}$
requires special investigation. In this region
$K_{1}(\mu_{G}\rho_{1})K_{1}(\mu_{G}\rho_{2})$ is exponentially
small, which is related to the exponential decay of the
correlation function of perturbative gluons. (Such a decay with
the correlation radius
$R_{c}\sim 0.2-0.3$f is suggested by the lattice studies, for the
review see [9]. We remind that $K_{1}(x)$ in the kernel ${\cal K}$
comes from the gradient of the gluon correlation function
$\propto K_{0}(x)$.)
. Then, at large $r$ our
Eq.~(\ref{eq:2.6}) takes the form
\arr
\sigma_{n+1}(r)=
{3\alpha_{S} \over \pi^{3}} \int d^{2}\vec{\rho}_{1}\,\,
\mu_{G}^{2}
K_{1}(\mu_{G}\rho_{1})^{2}
[\sigma_{n}(\rho_{1}) +\sigma_{n}(\rho_{2})-\sigma_{n}(r)]\, ,
\label{eq:4.3}
\endarr
which has a continuum of
solutions with the large-$r$ behavior of the form
\beq
E_{+}(\beta,\xi,r)= [a(\beta)+\exp(i\beta\mu_{G}r)]
\exp[\delta(\beta)\xi]\, .
\label{eq:4.4}
\endeq

The scaling limit of Eq.~(\ref{eq:2.6}) does not have localized
solutions, see eigenfunctions (\ref{eq:3.2}). The node-free
localized solutions of (\ref{eq:4.3}) which vanish at large $r$,
are excluded since for such a solution the l.h.s.
of Eq.~(\ref{eq:4.3})
vanishes at $r\rightarrow \infty$, as well as $\sigma(r)$ and
$\sigma(\rho_{2})$ in the integrand, and one is left with the
$r$-independent contribution from $\sigma(\rho_{1})$. This shows
that (\ref{eq:4.4}) gives a complete continuum set of solutions,
and in this limit our generalized BFKL equation generates a cut
in the complex $j$-plane.

When continued to small $r$
through the region of $r\sim R_{c}$, the plane waves
in the linear-$r$ space transform into the plane waves in the
$\log r^{2}$-space (times the overall factor $r$),
$E(i\nu,\xi,r)=r\exp(i\nu\log r^{2})$ of Eq.~(\ref{eq:3.2}),
so that $\delta(\beta)=\Delta(i\nu)$.
Evidently, the two real and the node-free
solutions with $\nu=0$ and $\beta=0$ must match each
other, so that the rightmost $j$-plane singularity with the
intercept $\Delta_{\Pom}$ Eq.~(\ref{eq:3.4})
must correspond to $\sigma_{\Pom}(r)\propto r$ at $r \ll R_{c}$ and
$\sigma_{\Pom}(r)\propto const$ at $r \gg R_{c}$.
We wish to emphasize that although such a $\sigma_{\Pom}(r)$ extends
to large $r$, Eq.~(\ref{eq:4.3}) makes it obvious that
the intercept $\Delta_{\Pom}$ is controlled by the behavior of
$\sigma_{\Pom}(r)$ at $r\sim R_{c}$.

The intercept
$\delta(\beta)$ will have a maximum at $\beta=0$.
Then, repeating the derivation of the Green's function
(\ref{eq:3.6}) one can easily show that at large $r,r'$
the Green's function of Eq.~(\ref{eq:4.3}) has the component
\beq
K(\xi,r.r')\propto {\exp(\Delta_{\Pom}\xi)\over \sqrt{\xi}}
\exp\left(-{\mu_{G}^{2}(r-r')^{2} \over 2\xi\delta''(0)}\right) \, ,
\label{eq:4.5}
\endeq
which is reminiscent of the familiar multiperipheral diffusion.
Indeed, at large $r\gg 1/\mu_{G}$ a sort of the additive quark
model is recovered, in which the (anti)quark of the dipole
developes its own perturbative gluonic cloud, and the
quark-quark scattering will be
described by the multiperipheral exchange of massive vector
mesons. The Green's function (\ref{eq:4.5}) shows the emergence
of the Regge growth of the interaction radius in this limit.
It also shows that the large-$\xi$ behavior of solutions of
our generalized BFKL equation in this region will be similar to
solutions of the Schr\"odinger equation
\beq
\left[-{\delta''(0)\over 2\mu_{G}^{2}}
{\partial^{2} \over \partial r^{2} }
+U(r)\right]\Phi= -{\partial \over \partial \xi} \Phi
=\epsilon \Phi
\label{eq:4.6}
\endeq
with the potential $U(r)=-\delta(0)$ which is flat at $r> R_{c}$.

The second interesting case is of $R_{c}\rightarrow \infty$,
i.e., the case of massless gluons, $\mu_{G}\rightarrow 0$,
but with the running coupling which freezes,
$\alpha_{S}(r)=\alpha_{S}^{(fr)}=\alpha_{S}(R_{f})$,
at finite $r \geq R_{f}$. In this case the scaling invariance
of the kernel ${\cal K}$
is restored on the infinite semiaxis $\log r > \log R_{c}$, where the
corresponding eigenfunctions $E_{+}(i\nu,\xi,r)$
are again essentially identical to the BFKL set
(\ref{eq:3.2}), the spectrum of eigenvalues will evidently
be continuous, the intercept $\Delta_{\Pom}$ will be given by
the BFKL formula with $\alpha_{S}=\alpha_{S}^{(fr)}$,
and the partial waves will have the
cut in the complex
$j$-plane identical to that for the original BFKL
equation. Indeed, the corresponding `Schr\"odinger' equation
(\ref{eq:4.2}) has the continuous spectrum starting with the minimum
energy $\epsilon = -\Delta(0)$, and the exact shape of the potential
at $z < z_{f}$ is not important here.
 Here we agree with Ross et al. [10,11] and
disagree with Lipatov [3], who concluded
that introduction of the running coupling leads
to the discret spectrum of eigenvalues.

The continuation of the BFKL solutions
$(\ref{eq:3.2})$ to the region of $z<z_{f}=\log R_{f}^{2}$ poses
no problems. Let us start with the
quasiclassical situation when $\alpha^{(fr)} \ll 1$, so that
$\alpha_{S}(z)=\alpha_{S}(R_{f})
\left[1+(\beta_{0}/4\pi)\alpha_{S}(R_{f})(z_{f}-z)\right]^{-1}$
is a slow function of $z$. In this case
the slowly varying running coupling can be factored out from
the integrand of Eqs.~(\ref{eq:2.6},\ref{eq:2.7}), which suggests
that
Eq.(\ref{eq:4.2}) will take the form of the 'Schr\"odinger'
equation for a particle with the slowly varying mass
\beq
M\approx {1\over \Delta''(0)}\cdot
{\alpha_{S}(R_{f})\over \alpha_{S}(r)}
\label{eq:4.7}
\endeq
in the
slowly varying and monotonically decreasing
potential
\beq
V(z) \approx -\Delta_{\Pom}{\alpha_{S}(r)\over \alpha_{S}(R_{f})}\, ,
\label{eq:4.8}
\endeq
which is flat, $V(z)=-\Delta_{\Pom}$, at $ z> z_{f}$.
Evidently, the solutions
$E_{+}(i\nu,\xi,r)$ with $\Delta(i\nu) > 0$ will have the
underbarrier decrease at
$z \rightarrow -\infty$, whereas the solutions with
$\Delta(i\nu) <0$ will be continued to $z \ll z_{f}$
as plane waves. Consequently, the eigenfunction $\sigma_{\Pom}(r)$
for the rightmost singularity will decrease at $r\rightarrow 0$
faster than the $r^{1}$ solution for the fixed-$\alpha_{S}$
BFKL scaling regime. The case
of large values of the frozen coupling $\alpha_{S}^{(fr)}$
must be qualitatively the same.

The  realistic case of the interest is
a combination of the two previous cases with $R_{f}\sim R_{c}$.
It is convenient
to start from the region of $r >R_{c},R_{f}$. The only change
from Eq.~(\ref{eq:4.3}) will be that the running coupling
$\alpha_{S}(\rho_{1})$ must be absorbed into the integrand.
We have a continuum of solutions of the form
(\ref{eq:4.4}) with the spectral parameter $-\infty <
\beta  < +\infty$, and by the same consideration as above
we can exclude the node-free localized
solutions with $\sigma(r)$ vanishing at large $r$.
The rightmost singularity in the
$j$-plane will have the dipole cross section
$\sigma_{\Pom}(r) = const$ vs. $r$ at large $r$. Its intercept
will be controlled by the behavior of $\sigma_{\Pom}(r)$ at
$r\sim R_{c}$.
The continuation of solutions (\ref{eq:4.4}) into the region of
$r < R_{c},R_{f}$ is not any different from that discussed above.
This shows, that our generalized BFKL equation
produces a cut in the complex $j$-plane.

The above
elimination of localized solutions and of the discret spectrum
of the pomeron is rigorous in the framework of our minimal
infrared regularization. Technically, it is based on the kernel
${\cal K}$ being finite at $r\rightarrow \infty$, see
Eq.~(\ref{eq:4.3}), which gives rise to $\sigma(r) \sim const$ at
$r> R_{c}$. The interaction
picture which emerges at $r> R_{c}$ has much intuitive appeal:
each well separated quark of the beam (target) dipole developes
the perturbative gluonic cloud of its own and a sort of the
additive quark model is recovered. The discret spectrum of the
pomeron comes along with the localized solutions
$\sigma(r)\rightarrow 0$ at $r\rightarrow \infty$. At first sight,
such a vanishing of the perturbative
total cross section for colour dipoles of large size
looks quite unphysical, but we wish to present the qualitative
arguments in favor of such a possibility.

The plausible scenario for the discret spectrum of the pomeron is
as follows: Our derivation clearly shows that the
kernel ${\cal K}$ is proportional to the probability
of radiation of perturbative gluons. When the quarks of the colour
dipole are a distance $r\gg R_{c}$ apart, the nonperturbative
colour fileds stretched between the quarks may strongly modify
the vacuum and suppress the perturbative gluonic fluctuations
on the nonperturbative background in the vicinity of
quarks. This can be modelled by the decrease of the
effective (nonlocal) perturbative coupling $\alpha_{S}(r,\rho)$
and/or the increase of $\mu_{G}$ with increasing $r$, resulting
in the decrease of the kernel ${\cal K}$ with increasing $r$.
In terms of the Schr\"odinger equation (\ref{eq:4.6}) this amounts       for
to an increase of $U(r)$ towards large $r$. (One can draw a useful
analogy with the asymptotic freedom decrease of
$\alpha_{S}(r)$ and the related increase of the $V(z)$
Eq.~(\ref{eq:4.8}) in the Schr\"odinger equation (\ref{eq:4.2}).)
Evidently, the energy $\epsilon$ of the lowest
state will be higher than the bottom of the potential well, so
that the intercept $\Delta_{\Pom}$ of the rightmost singularity
will be lowered compared to the case of the minimal regularization.
If with the increasing $r$ the potential $U(r)$ flattens at the
still negative value $U(\infty)=-\Delta_{c}$, then the Schr\"odinger
equation (\ref{eq:4.6}) will have the continuous spectrum starting
with $\epsilon = -\Delta_{c}$ and in the complex
$j$-plane there will be a cut from $j=1+\Delta_{c}$ to $j=-\infty$
and certain number of isolated poles to the right of the cut.
If $U(\infty) >0$, then the cut in the complex $j$-plane will
start at $j=1$.
Ross and Hancock [11] find a discrete spectrum in the model
with a very different infrared cutoff of the original BFKL
equation in the momentum representation. We have checked that
even a very abrupt cutoff of $r>R_{conf}\sim 1.5$f has a negligible,
less than one per cent, effect on $\Delta_{\Pom}$.

%-----------------------------------------------

%----------------------------     Section 5

\section{The evaluation of $\Delta_{\Pom}$ and the pomeron dipole
cross section}

We study the large-$\xi$ behavior of
numerical solutions of Eq.~(\ref{eq:2.7}) and verify that
$\sigma(\xi,r)$ has the same asymptotic behavior $\sigma(\xi,r)
\Longrightarrow \sigma_{\Pom}(r)\exp(\Delta_{\Pom}\xi)$
irrespective
of the boundary condition at $\xi=0$. Namely, we compute
$\Delta_{eff}(\xi,r)=\partial \log\sigma(\xi,r)/\partial \xi$
and check that at large $\xi$ the effective intercept
$\Delta_{eff}(\xi,r)$ tends to the same limiting value
$\Delta_{\Pom}$ for all $r$.
In Fig.1 we present the eigenfuction $\sigma_{\Pom}(r)$
for few values of $\mu_{G}$ (we keep $\alpha_{S}^{(fr)}=0.8$).
As we have
anticipated above, $\sigma_{\Pom}(r)$ flattens at large $r$
and decreases towards small $r$ faster than the BFKL solution
$\propto r^{1}$  of Eq.~(\ref{eq:3.7}), but slower than
$\propto r^{2}$.

The corresponding intercepts are shown in Fig.2. Our
$\Delta_{\Pom}$
is significantly
smaller than the BFKL result (\ref{eq:3.4}) and
smaller than the Collins-Kwiecinski (CK) lower bound [12]
\beq
\Delta_{\Pom} > {3.6 \over \pi}\alpha_{S}^{(fr)}\,.
\label{eq:5.1}
\endeq
This bound was derived [12] using the
sharp infrared cutoff of $k^{2}>k_{0}^{2}$. For closer connection
with the CK analysis in Eq.~(\ref{eq:5.1}) we substitute
$\alpha_{S}^{(fr)}$ for ${\rm min}\{\alpha_{S}^{(fr)},
\alpha_{S}(\mu_{G}^{2})\}$. The significance of the CK lower
bound is not clear to us, since the CK cutoff breaks the initial
symmetry of the BFKL equation, does not
give an identical cutoff of all gluon propagators and is suspect
from the point of view of gauge invariance ([13], for the
similar criticism of the cutoff [12] see also Collins and
Landshoff [8]; this criticism also applies to the cutoff used in [11]
).)

Above we have used very simple
modelling of the nonperturbative effects in the infrared region
in terms of the gluon correlation radius $R_{c}$ and the freezing
coupling $\alpha_{S}(r)$. Because we have checked that
$\Delta_{\Pom}$ only weakly depends on the contribution from
$r\gsim $1-2f, we conclude
that with the realistic values of $R_{c}\sim $0.2-0.3f,
suggested by the lattice studies, the intercept $\Delta_{\Pom}$
for the exchange by perturbative gluons will be
significantly higher
than the effective intercept $\Delta_{\Pom}(hN) \sim 0.1$ as
given by the phenomenology of hadronic scattering [14,15].
The plausible scenario, suggested by the observed slow rise
of the hadronic cross sections is that the exchange by perturbative
gluons is only a small part of $\sigma_{tot}(hN)$ at moderate
energies ([5], for the early discussion of this scenario see [14]).
The detailed phenomenology of the hadronic cross sections and
of the diffractive deep inelastic scattering in the framework
of our generalized BFKL equation will be presented elsewhere,
we only notice that the choice of $\mu_{G}=0.75$ gives a
consistent description of the energy dependence of total cross
section for the hadronic scattering and the real photoproduction,
and of the small-$x$ structure functions [16].

In the above discussion we have suppressed the target size
variables $r_{B}$ but, evidently,
it is the generalized energy-dependent dipole-dipole cross section
$\sigma(\xi,\vec{r}_{A},\vec{r}_{B})$ which
emerges as the fundamental
quantitiy of the lightcone $s$-channel approach to the diffractive
scattering. The lowest-order dipole-dipole cross section has an
obvious beam-target symmetry property
$\sigma_{0}(\vec{r}_{A},\vec{r}_{B})=
\sigma_{0}(\vec{r}_{B},\vec{r}_{A})$. This beam-target symmetry is
but a requirement of the Lorentz-invariance, one has to
have it extended to all energies, and it
emerges as an important constraint on the admissible boundary
condition for the BFKL equation.
In the derivation of our generalized
BFKL equation we have treated the $s$-channel gluon $g_{s}$ of
Fig.~3 as belonging to the beam-dipole $A$. The gluon-induced
correction to the total cross section was reinterpreted in
terms of the generalized dipole cross section
$\sigma(\xi,\vec{r}_{A},\vec{r}_{B})$, see Eq.~(\ref{eq:2.5}).
Alternatively, we could have treated the same $s$-channel gluon
$g_{s}$ as belonging to the target-dipole, and the result must
have been the same. The two descriptions differ in that in the
former the perturbative $t$-cahnnel gluons $g_{1}$ and $g_{1'}$
enter the kernel ${\cal K_{A}}$, which we supply with the
subscript $A$ as it acts on the beam variable $r_{A}$ of the
dipole-dipole cross section $\sigma(\xi,r_{A},r_{B})$,
whereas the gluons $g_{2}$ and $g_{2'}$ are the exchanged gluons
in the dipole-dipole cross section (\ref{eq:1.1}). In the latter
description $g_{2},g_{2'}$ enter the kernel ${\cal K_{B}}$ which
now acts on the target variable $r_{B}$ of the dipole-dipole
cross section $\sigma(\xi,\vec{r}_{A},\vec{r}_{B})$, and
$g_{1},g_{1'}$ will become the exchanged gluons in the input
dipole-dipole cross section. The beam-target symmetry constraint
essentially implies, that the boundary condition for the BFKL
evolution must be calculable within the same perturbation theory as
the one used to construct the generalized BFKL kernel ${\cal K}$
Eq.~(\ref{eq:2.6}). (To this end we remind, that the $K_{1}(x)$
in the kernel ${\cal K}$ is precisely the derivative
of the gluon propagator (correlation function) $K_{0}(x)$.) This
kernel-cross section
relationship is crucial for having the beam-target symmetry, which
will be violated with the arbitrary choice of the boundary
condition for the dipole-dipole cross section
$\sigma(\xi=0,\vec{r}_{A},\vec{r}_{B})$.

We conclude with the comment, that for the rightmost singularity
in the $j$-plane
the beam-target symmetric dipole-dipole total cross section will
have the factorized form
\beq
\sigma_{\Pom}(\xi,\vec{r}_{A},\vec{r}_{B})=
\sigma_{\Pom}(r_{A})\sigma_{\Pom}(r_{B})\exp(\xi\Delta_{\Pom})\, .
\label{eq:5.2}
\endeq
Remarkably, by virtue of Eq.~(\ref{eq:1.2}) this implies that
at asymptotic energies the BFKL pomeron gives rise to the
factorization of total cross sections. However, this factorization
will be broken by the unitarization corrections needed to tame
too rapid a rise of the bare pomeron cross section.

\section{Conclusions}

The purpose of this paper has been to understand the spectrum of
eigenvalues of the generalized BFKL equation [4,5] for the dipole
total cross section in the
realistic case of a finite correlation radius $R_{c}$ for
perturbative gluons and of the freezing strong coupling
$\alpha_{S}(r)$.
The advantage of our equation (\ref{eq:2.6},\ref{eq:2.7}) is an
easy introduction of the gluon correlation radius $R_{c}$ in a
manner which is consistent with gauge invariance constraints.

We have shown that our generalized BFKL equation
(\ref{eq:2.6},\ref{eq:2.7}) has the continuous spectrum, which
corresponds to the QCD pomeron described by the cut in the
complex $j$-plane. (Under certain conditions, the nonperturbative
effects may produce poles in the $j$-plane, though, and we
commented on these conditions.) We have determined
the dipole cross section $\sigma_{\Pom}(r)$
for the rightmost singularity in the $j$-plane
and found the corresponding intercept $\Delta_{\Pom}$.
It is much smaller than given by the BFKL formula (\ref{eq:3.4})
and even smaller than the lower bound cited in [12].
Still, with the realistic infrared regularization we find
$\Delta_{\Pom}$ which is substantially larger than the
phenomenological value $\Delta_{\Pom}(hN) \sim 0.1$ suggested
by the observed energy-dependence of hadronic cross sections.
The plausible solution is that the exchange by perturbative
gluons contributes only a small part of the hadronic total
cross sections and more work in this direction is needed.
We have formulated the beam-target symmetry as the consistency
constraint  which must be satisfied by the boundary condition
for the BFKL equation for total cross section. We have shown
that the bare BFKL pomeron gives the factorizing asymptotic
total cross section.
\bigskip\\
{\bf Acknowledgements}: B.G.Z. and V.R.Z. are grateful to
J.Speth for the hospitality at IKP, KFA J\"ulich.
\pagebreak

\pagebreak
\begin{itemize}
\item[Fig.1 - ]
The pomeron dipole cross section $\sigma_{\Pom}(r)$ for different
values of $\mu_{G}$. The straight lines show the $r^{1}$ and $r^{2}$
behavior.

\item[Fig.2 - ]
The intercept $\Delta_{\Pom}$ for $\mu_{G}=0.3,\,0.5,\,0.75,\,1.0\,$
GeV. The solid curve shows the Collins-Kwiecinski
 lower bound (\ref{eq:5.1}).

\item[Fig.3 - ]
One of diagrams of the driving term of the rising total cross
section.
\end{itemize}
\end{document}